\documentclass[twocolumn,showpacs,preprintnumbers,amsmath,amssymb]{revtex4}

\usepackage{epsf}

\usepackage{graphicx}
\usepackage{dcolumn}
\usepackage{bm}

\newcommand{\bd     }{\begin{displaymath}}
\newcommand{\ed     }{\end{displaymath}}

\newcommand{\s      }{\sigma}

\newcommand{\bra    }{\langle}
\newcommand{\ket    }{\rangle}
\newcommand{\Bra    }{\left\langle}
\newcommand{\Ket    }{\right\rangle}

\newcommand{\bh     }{\mbox{\boldmath$h$}}

\newcommand{\bJ     }{\mbox{\boldmath$J$}}
\newcommand{\bc     }{\mbox{\boldmath$c$}}

\newcommand{\bsigma }{\mbox{\boldmath$\sigma$}}

\newcommand{\btau   }{\mbox{\boldmath$\tau$}}

\newcommand{\be     }{\mbox{\boldmath$e$}}

\newcommand{\bs     }{\mbox{\boldmath$s$}}

\newcommand{\bx     }{\mbox{\boldmath$x$}}
\newcommand{\by     }{\mbox{\boldmath$y$}}

\usepackage{color}

\begin{document}

\title{On metastable configurations of small-world networks}

\author{R. Heylen, N.S.\@ Skantzos, J. Busquets Blanco and D. Boll\'e}
\affiliation{Instituut voor Theoretische Fysica, Celestijnenlaan 200D,
   Katholieke Universiteit Leuven, B-3001 Leuven, Belgium}
\email{rob.heylen@fys.kuleuven.be}


\begin{abstract}
We calculate the number of metastable configurations of Ising
small-world networks which are constructed upon superimposing
sparse Poisson random graphs onto a one-dimensional chain. Our
solution is based on replicated transfer-matrix techniques.
We examine the denegeracy of the ground state and we find
a jump in the entropy of metastable configurations exactly
at the crossover between the small-world and the Poisson random graph structures.
We also examine the difference in entropy between metastable 
and all possible configurations, for both
ferromagnetic and bond-disordered long-range couplings.
\end{abstract}

\maketitle

\section{Introduction}

Small-world systems exhibit remarkable cooperation, not found in
complex systems with e.g.\@ an Erd\"os-Renyi
structure. The origins of the appellation `small-world' can be
traced to the now famous experiment by the Harvard social
psychologist Stanley Milgram \cite{milgram}. The outcome of this
experiment pointed to the fact that the structure of many real
networks is such that distant nodes can in fact be connected via
long-range short-cuts. This architectural property leads to small
path lengths between any pair of nodes and thus enhances
information processing and cooperation.

The importance and ubiquitous nature of small-world structures in
complex networks received further attention by the seminal paper
of Watts and Strogatz \cite{wattsstrogatz} in which the authors
proposed the small-world structure as a way to interpolate between
so-called `regular' and  `random' networks. Surprisingly, the
simple small-world architecture can be found in many different
circumstances, ranging from linguistic, epidemic and social
networks to the world-wide-web. By now, a large body of work has
been devoted to the study of small-world networks, mainly
numerical, with emphasis e.g.\@ on biophysical
\cite{girvan,simingli,barabasi_nature} or social networks
\cite{newman_pnas} and to a lesser extent analytical
\cite{barrat,theodore}. For recent reviews in the area of
small-worlds see e.g.\@ the articles
\cite{albert-barabasi,newman_review}, or the books
\cite{bookwatts,bookdoro}.

From a statistical mechanical point of view such systems combine
two universality classes: a sparse `graph' structure, which is
superimposed upon a one-dimensional `ring'. Thus, every node on
the ring has a local neighborhood and a certain number of
long-range connections to distant parts of the chain. It was shown
in \cite{theodore} (and also in \cite{XYsmall} for the case of XY
spins) that this construction significantly enlarges the region in
parameter space where ferromagnetism occurs. In particular, it was
shown that the ferromagnetic-paramagnetic transition always occurs
at a finite temperature for any value of the average long-range
connectivity (however small). On a technical level, in evaluating
the relevant disorder-averaged free energy one is immediately
confronted with the problem of diagonalising a $2^n\times 2^n$
transfer matrix where $n$ represents the replica dimension.
Although for an infinite system size obtaining only the largest
eigenvalue suffices for the evaluation of the free energy, one can
in principle follow the systematic analysis of
\cite{diagonalisation} to derive the entire spectrum of
eigenvalues and thus evaluate e.g.\@ correlation functions.

Several important issues remain to be understood for small-world
systems. In this paper we evaluate the number of metastable
configurations, or more precisely, the number of equilibrium
configurations in which spins align to their local fields. With
this definition the energy of the system in a metastable state
cannot be decreased by flipping a single spin. Such
configurations can be e.g.\@ responsible for trapping the
microscopic update dynamics in locally stable states. Thus, from
an experimental point of view it is advantageous to know what is
the relevant size in phase space occupied by such states.

The computation of the number of metastable configurations in
small-world systems is generally an involved problem, both
analytically and numerically. Indicatively, on sparse random
graphs structures without the superimposed `ring' the evaluation of the
number of metastable configurations, or the so-called configurational
entropy, has only recently taken off
\cite{bergsellito,pagnani,lefevre} following the course of the
relevant analytic techniques (as e.g.\@ in
\cite{mezardparisi,mezardparisi0,leone}). In particular, using the
replica method, the solution of the ferromagnetic Poisson graph
has been studied in \cite{bergsellito} while with the cavity
method the authors of \cite{pagnani} examined the bond-disordered
Bethe lattice. These results agree well with results of numerical 
enumerations \cite{boettcher1,boettcher2} and also serve as
good limiting tests of our findings.

This paper is organized as follows: in Section \ref{sec:definitions} we
first define the small-world model. In Section \ref{sec:saddle-points} we
express the generating function of the system as a saddle-point
problem which we then evaluate in Section \ref{sec:RTM} using
replicated transfer matrix techniques and within the replica symmetric
approximation. In Section \ref{sec:results} we present our
results and finally we present some concluding remarks in Section
\ref{sec:conclusions}.

\section{The model}
\label{sec:definitions}

Our model describes a system of $N$ Ising spins
$\bsigma=(\sigma_1,\ldots,\sigma_N)$, with $\sigma_i\in\{-1,1\}$,
arranged on a one-dimensional lattice. There are two different
couplings in this system: firstly, nearest-neighbor interactions of
uniform strength $J_0$ and secondly, sparse long-range ones. To
model the latter we will assign the random variable $c_{ij}$ for
every pair of sites $(i,j)$ representing whether a connection
exists ($c_{ij}=1$) or not ($c_{ij}=0$). This variable will be taken for
all $i<j$ from the distribution
\begin{equation}
Q_c(c_{ij})=\frac{c}{N}\delta_{c_{ij},1}+\left(1-\frac{c}{N}\right)\delta_{c_{ij},0},\label{eq.dist}
\end{equation}
so that, on average, every site has $c$ long-range connections. In
the small-world context one takes $c$ to be a small number of
order $\mathcal{O}(1)$ while $c/N\to 0$. The bond-strength $J_{ij}$ of the
long-range coupling between any pair of spins $(i,j)$ (with $i<j$)
will be taken from the distribution
\begin{equation}
Q_J(J_{ij})=p\,\delta_{J_{ij},J}+(1-p)\,\delta_{J_{ij},-J},
\label{eq:bond_disorder}
\end{equation}
for some $J>0$, so that $p=1$ corresponds to a model with strictly
ferromagnetic interactions. To allow for detailed balance we  will
also consider absence of self-interactions and symmetry of the
connectivity matrix, namely $c_{ii}=0$, $c_{ij}=c_{ji}$ and
$J_{ij}=J_{ji}$. At thermal equilibrium the above system can be
described by the Hamiltonian:
\begin{equation}
\mathcal{H}(\bsigma)=-\frac12\sum_i \s_i\, h_i(\bsigma),
\label{eq:H}
\end{equation}
with the local fields defined as:
\begin{eqnarray*}
h_i(\bsigma)&=&
\sum_j\left[J_0(\delta_{j,i+1}+\delta_{j,i-1})+\frac{c_{ij}}{c}J_{ij}\right]\s_j.
\end{eqnarray*}
We now impose the condition  for metastability: in a similar
spirit as e.g.\@ in \cite{braymoore,gardner} we call a
configuration $\bsigma$ metastable if all spins align to their
local fields, i.e.\@
\begin{eqnarray*}
\prod_i \Theta(\s_ih_i(\bsigma)) = 1,
\end{eqnarray*}
where $\Theta(x)=1$ for $x\geq 0$ and $\Theta(x)=0$ otherwise.
Notice that we have taken $\Theta(0)=1$ which is dictated by the underlying
physics: for  spins which receive a zero local field, the
energetic cost of aligning to either of the two possible
directions is identical. A consequence of the above definition is
that any metastable configuration $\bsigma^{\text{MS}}$ is a local
minimum of the Hamiltonian:
\begin{equation*}
\forall \btau :
\frac{1}{2}\sum_i\vert\sigma_i^{\text{MS}}-\tau_i\vert=1
\Rightarrow \mathcal{H}(\btau)\geq\mathcal{H}(\bsigma^{\text{MS}}).
\end{equation*}
Since we are interested in evaluating the number of metastable
configurations we will define the following generating function:
\begin{eqnarray}
-\beta f&=& \lim_{N\to\infty}\frac1N\Bra \log
\sum_{\bsigma}e^{-\beta
\mathcal{H}(\bsigma)}\prod_{i=1}^N\Theta(\s_ih_i(\bsigma))\Ket_{\bc,\bJ}
\label{eq:generating_function}
\end{eqnarray}
where $\beta$ represents the inverse temperature and
$\bc=\{c_{ij}\}$, $\bJ=\{J_{ij}\}$. As in \cite{bergsellito,pagnani}, the entropy density $s_m$ 
of the metastable configurations 
can be evaluated from (\ref{eq:generating_function}) via $s_m=\beta\partial_\beta(\beta  f)-\beta f$.

\section{Saddle-point equations}
\label{sec:saddle-points}

To evaluate the disorder average in (\ref{eq:generating_function})
we begin by invoking the replica identity: $\bra \log
Z\ket=\lim_{n\to 0}\frac1n\log \bra Z^n\ket$. As the disorder
variables $\{c_{ij},J_{ij}\}$ lie within the \mbox{$\Theta$-function} we
insert the following unity into our expression
\begin{equation}
1 = \int\prod_{i,\alpha} dh_i^\alpha\
\delta[h_i^\alpha-h_i(\bsigma^\alpha)],
\label{eq:field_def}
\end{equation}
with $\bsigma^{\alpha}=(\sigma_1^\alpha, \cdots, \sigma_N^\alpha)$
and $\alpha=1,\ldots,n$.  This allows us to conveniently re-locate
$c_{ij},J_{ij}$ into exponents where averages can be taken more
easily, namely
\begin{eqnarray}
\lefteqn{-\beta f=\lim_{N\to\infty}\lim_{n\to
0}\frac{1}{Nn}\log\int
\{dh_i^\alpha\,d\hat{h}_i^\alpha\}e^{-i\sum_{i,\alpha}h_i^\alpha\hat{h}_i^\alpha}\nonumber}
\\
& & \times\,\sum_{\{\sigma_i^\alpha\}}
\prod_{i,\alpha}\left[e^{iJ_0\hat{h}_i^\alpha(\s_{i+1}^\alpha+\s_{i-1}^\alpha)+
\frac12\beta\s_i^\alpha h_i^\alpha}\Theta(\s_i^\alpha
h_i^\alpha)\right]\nonumber
\\
& & \times\,\Bra
e^{i\sum_{i,\alpha}\hat{h}_i^\alpha\sum_j\frac{c_{ij}}{c}J_{ij}\s_j^\alpha}\Ket_{\bc,\bJ},
\label{eq:f_inter}
\end{eqnarray}
where
$\{dh_i^\alpha\,d\hat{h}_i^\alpha\}=\prod_{i,\alpha}(2\pi)^{-1}
dh_i^\alpha\,d\hat{h}_i^\alpha$. Let us concentrate on the last
line of the above which contains the disorder. After symmetrizing
with respect to the sites $i<j$ it leads for $N\to\infty$ to:
\begin{eqnarray}
\lefteqn{\Bra
e^{i\sum_{i,\alpha}\hat{h}_i^\alpha\sum_j\frac{c_{ij}}{c}J_{ij}\s_j^\alpha}\Ket_{\bc,\bJ}}
\nonumber
\\
&=& \exp\left[\frac{c}{2N}\sum_{ij}\left(\Bra
e^{i\frac{\mathcal{J}}{c}\sum_\alpha (\hat{h}_i^\alpha\sigma_j^\alpha+
\hat{h}_j^\alpha\sigma_i^\alpha)}\Ket_{\mathcal{J}}-1\right)\right]
\label{eq:dilav}
\end{eqnarray}
where $\bra \cdots\ket_{\mathcal{J}}$ denotes an average over the
binary random variable $\mathcal{J}$ taken from the distribution
$Q_J(\cdot)$, eq.\@ (\ref{eq:bond_disorder}). We have used the fact
that $c/N\to 0$ to recast the result of averaging over
$\{c_{ij}\}$ into an exponential form. Note now that upon
inserting the unities $1=\sum_{\bsigma}\delta_{\bsigma,\bsigma_i}$
and $1=\sum_{\btau}\delta_{\btau,\bsigma_j}$ where $\bsigma,\btau$
are auxiliary vectors in replica space (and we have denoted
$\bsigma_i=(\sigma_i^1,\ldots,\sigma_i^n)$), one has effectively
created an order function $P_{\mathcal{J}}(\bsigma,\btau)$. As usual it
can be inserted into
our generating function via:
\begin{eqnarray}
\lefteqn{1=\int\prod_{\bsigma\btau}\prod_{\mathcal{J}=\pm
J}dP_\mathcal{J}(\bsigma,\btau)\nonumber}
\\
&&
\times\,\delta\left[P_{\mathcal{J}}(\bsigma,\btau)-\frac{1}{N}\sum_i\delta_{\bsigma,\bsigma_i}\,
e^{i\frac{\mathcal{J}}{c}\hat{\bh}_i\cdot\btau}\right].
\label{eq:orderfunction_def}
\end{eqnarray}

As in \cite{topologies}, to understand the physical meaning of the
above order  function one needs to add a generating term in the
replicated Hamiltonian
$\sum_{\alpha}\mathcal{H}(\bsigma^\alpha)\to
\sum_\alpha\mathcal{H}(\bsigma^\alpha)+\eta
P_\mathcal{J}(\bsigma,\btau)$ and take the derivative $\partial
f/\partial \eta|_{\eta=0}$ in (\ref{eq:generating_function}). One
then sees that upon introducing the identities (\ref{eq:field_def})
the order function becomes the distribution of replicated spins
with one connection removed (equivalently, it becomes the
distribution of replicated `cavity' spins).

We now aim to eliminate from our expressions the set of fields
$\{h_i^\alpha,\hat{h}_i^\alpha\}$.
This can be done by replacing the delta function
in (\ref{eq:orderfunction_def}) by its Fourier representation (for details see appendix \ref{sec:ap1}).
As an end result we obtain an extremisation problem over the density
$P_\mathcal{J}(\bsigma,\btau)$ expressed in terms of a trace over a transfer function, namely
\begin{widetext}
\begin{equation}
-\beta f= \lim_{n\to 0}\frac{1}{n}\underset{P}{\rm Extr}
\left[-\frac{c}{2}\sum_{\bsigma\btau}\Bra
P_\mathcal{J}(\bsigma,\btau)P_\mathcal{J}(\btau,\bsigma)\Ket_\mathcal{J}
+\lim_{N\to\infty}\frac1N\log\,{\rm tr}\,(T^N[P])\right],
\label{eq:extr}
\end{equation}
\end{widetext}
where $\mathcal{J}\in\{-J,J\}$ and with the abbreviation ${\rm
tr}(T^N[P])={\rm tr}(T^N[P_{J},P_{-J}])$. The order functions
$P_{\pm J}$ are to be evaluated self-consistently from
\begin{eqnarray}
\lefteqn{P_\mathcal{J}(\bsigma,\btau)\propto
\frac{1}{N}\sum_{j=1}^N\sum_{\bsigma_1\cdots\bsigma_N}\delta_{\bsigma_j,\bsigma}\nonumber}
\\
&\times& \
F^{(\mathcal{J})}_{\bsigma_{j-1},\bsigma_j,\bsigma_{j+1}}[P,\bsigma,\btau]\prod_{i\neq
j} T_{\bsigma_{i-1},\bsigma_i,\bsigma_{i+1}}[P] \label{eq:Psigmatau}
\end{eqnarray}
We have absorbed the normalisation constant
$\sum_{\{\sigma^\alpha_i\}} \prod_i
T_{\bsigma_{i-1},\bsigma_i,\bsigma_{i+1}}[P]$ in the
proportionality symbol. The traces in (\ref{eq:extr}) and
(\ref{eq:Psigmatau}) involve correlations between next-nearest
neighbors and can be evaluated in a spirit similar to the transfer
matrix technique. The relevant tensor is defined over a $2^n\times
2^n\times 2^n$ space with elements
\begin{eqnarray}
\lefteqn{T_{\bsigma_{i-1},\bsigma_i,\bsigma_{i+1}}[P]=\sum_{\mu\geq
0}\frac{e^{-c}c^\mu}{\mu!} \sum_{\btau_1\cdots\btau_\mu} }
\label{eq:matrix_Ta}
\\
& & \times\,\Bra[\prod_{\nu\leq \mu}
P_{\mathcal{J}_\nu}(\btau_\nu,\bsigma_i)]\prod_\alpha
R_i^\alpha(\sum_{\nu\leq \mu}\mathcal{J}_\nu\tau^\alpha_\nu)\Ket_{\mathcal{J}_1\cdots\mathcal{J}_\mu}, \nonumber
\end{eqnarray}
with the convention that the $\mu=0$ term of the above equals $e^{-c}\prod_\alpha R_i^\alpha(0)$. The quantity $F$ that specifies our order function is
\begin{eqnarray}
\lefteqn{F^{(\mathcal{J})}_{\bsigma_{i-1},\bsigma_i,\bsigma_{i+1}}[P,\bsigma,\btau]=\sum_{\mu\geq
0}\frac{e^{-c}c^\mu}{\mu!}
  \label{eq:matrix_Fa}
\sum_{\btau_1\cdots\btau_\mu}
  }
\\
& & \times \Bra[\prod_{\nu\leq \mu}
P_{\mathcal{J}_\nu}(\btau_\nu,\bsigma)]\prod_\alpha
R_i^\alpha(\sum_{\nu\leq
\mu}\mathcal{J}_\nu\tau^\alpha_\nu+\mathcal{J}\tau^\alpha)\Ket_{\mathcal{J}_1\cdots\mathcal{J}_\mu}.
\nonumber
\end{eqnarray}
We have used the abbreviation
\begin{eqnarray}
\lefteqn{R_i^\alpha(x) =} \label{eq:R}
\\
& &
e^{\frac{1}{2}\beta\sigma_i^{\alpha}(J_0(\s_{i+1}^\alpha+\s_{i-1}^\alpha)+\frac{x}{c}
)}
\Theta[\sigma_i^{\alpha}(J_0(\s_{i+1}^\alpha+\s_{i-1}^\alpha)+\frac{x}{c}
)].\nonumber
\end{eqnarray}
Note that due to our symmetrization with respect to site indices in
(\ref{eq:dilav}) we have ended up with a symmetric quantity, namely
$T_{L\bx_i}=T_{\bx_i}$ where $L$ is the 3$\times$3 matrix $L_{\ell k}=\delta_{\ell+k,4}$ and $\bx_i=
(\bsigma_{i-1},\bsigma_i,\bsigma_{i+1})$.

The structure of the function $R_i^\alpha(x)$ indicates that the
input $x$ is related to the long-range field received by a site
$i$. The tensors $T$ and $F$ differ only in their input to this
function. Since this is proportional to $\pm J/c$ we understand
that it is related to the `effective' (or, `cavity') and true
local-field, respectively.

\section{Replicated transfer matrix analysis}
\label{sec:RTM}

To interpret the spin summations as matrix multiplications in
equations (\ref{eq:extr}) and (\ref{eq:Psigmatau}) we need
to transform our variables such that the traces in these equations
involve nearest-neighbor correlations only. This can be done in several
ways. For instance, let us introduce the auxiliary spins:
\begin{equation}
\bs_i^\alpha\equiv(s_i^{\alpha,1},s_i^{\alpha,2})=(\s_i^\alpha,\s^\alpha_{i+1}).
\label{eq:transform}
\end{equation}
To suppress the replica index above, we will occasionally use
the more compact notation
$\bs_i=(\bs^{(1)}_i,\bs^{(2)}_i)=(\bsigma_{i},\bsigma_{i+1})$ with
vectors now defined in replica space. With the above, we can now transform 
$T_{\bsigma_{i-1},\bsigma_i,\bsigma_{i+1}}$ into 
$T_{\bs_{i-1},\bs_i}$ and in particular
\begin{eqnarray}
\lefteqn{\hspace{-4mm}T_{\bs_{i-1},\bs_i}[P]=
\delta_{\bs_{i-1}^{(2)},\bs_{i}^{(1)}}\sum_{\mu\geq
0}\frac{e^{-c}c^\mu}{\mu!}\sum_{\btau_1\cdots\btau_\mu} \nonumber}
\\
& & \times\Bra[\prod_{\nu\leq \mu}
P_{\mathcal{J}_\nu}(\btau_\nu,\be\cdot\bs_i)]\prod_\alpha
\hat{R}_i^\alpha(\sum_{\nu\leq \mu}\mathcal{J}_\nu\tau^\alpha_\nu)\Ket_{\mathcal{J}_1\cdots\mathcal{J}_\mu}
\label{eq:matrix_T}
\end{eqnarray}
with $\be=(1,0)$. Similarly for $F$:
\begin{eqnarray}
\lefteqn{F^{(\mathcal{J})}_{\bs_{i-1},\bs_i}[P,\bsigma,\btau]=
\delta_{\bs_{i-1}^{(2)},\bs_{i}^{(1)}}\sum_{\mu\geq
0}\frac{e^{-c}c^\mu}{\mu!} \sum_{\btau_1\cdots\btau_\mu}}
\nonumber \label{eq:matrix_F}
\\
& & \times\Bra[\prod_{\nu\leq \mu}
P_{\mathcal{J}_\nu}(\btau_\nu,\bsigma)]\prod_\alpha \hat{R}_i^\alpha(\sum_{\nu\leq
\mu}\mathcal{J}_\nu\tau^\alpha_\nu+\mathcal{J}\tau^\alpha)\Ket_{\mathcal{J}_1\cdots\mathcal{J}_\mu}. \nonumber
\end{eqnarray}
The Kronecker deltas on the right-hand side of the above
expressions impose the transformation (\ref{eq:transform}). Similarly
to equation (\ref{eq:R}) we now have
\begin{eqnarray}
\hat{R}_i^\alpha(x)= e^{\frac12\beta\bs_i^\alpha\cdot(
J_0\bs_{i-1}^\alpha+ \frac{x }{c}\be ) }\
\Theta\left[\bs_i^\alpha\cdot( J_0\bs_{i-1}^\alpha+ \frac{x
}{c}\be )\right]. \nonumber
\end{eqnarray}
This transfer matrix is also symmetric. In particular, it obeys $T_{\hat{L}\by_i}=T_{\by_i}$ where
$\hat{L}$ is the 4$\times$4 matrix $\hat{L}_{\ell k}=\delta_{\ell+k,5}$ (the so-called Dirac matrix $E_{11}$) and $\by_i=(\bs_i,\bs_{i+1})$.
With the above definitions we can now write the self-consistent
equation (\ref{eq:Psigmatau}) in a transparent way:
\begin{equation}
P_\mathcal{J}(\bsigma,\btau)=\sum_{\bsigma'}\frac{{\rm
tr}\left(Q^{(\mathcal{J})}[P,\bsigma',\bsigma,\btau]\,
T^{N-1}[P]\right)} {{\rm tr}\left(T^{N}[P]\right)}, \label{eq:P_tr}
\end{equation}
with the auxiliary matrix
\begin{equation}
Q^{(\mathcal{J})}_{\bs_a,\bs_b}[P,\bsigma',\bsigma,\btau]\equiv
F^{(\mathcal{J})}_{\bs_a,\bs_b}[P,\bsigma,\btau]\ 
\delta\left[{\bs_a^{(1)} \choose \bs_a^{(2)}}-{\bsigma'
\choose \bsigma}\right] \label{eq:Q}
\end{equation}
To proceed with the evaluation of
the traces involved in (\ref{eq:extr}) and (\ref{eq:P_tr}) we now
aim at diagonalizing the transfer matrix
$T$. Our analysis hereafter will follow closely
\cite{theodore}. To this end, let us consider the eigenvector equation
corresponding to the largest eigenvalue $\lambda_0(n)$, namely
\begin{eqnarray}
\sum_{\bs'}T_{\bs\bs'}[P]\ u(\bs')&=&\lambda_{0}(n)\ u(\bs).
\label{eq:right}
\end{eqnarray}
Note that we have only defined a `right' eigenvector. It is
sufficient due the symmetry of our transfer matrix. Next, using
(\ref{eq:Q}) and (\ref{eq:right})  we can rewrite the
self-consistent equation (\ref{eq:P_tr}) in terms of $u(\bs)$
\begin{eqnarray}
\lefteqn{P_\mathcal{J}(\bsigma,\btau)=\label{eq:P_decomp}}
\\
&&
\frac{\sum_{\bsigma'\bs\bs'}F_{\bs\bs'}^{(\mathcal{J})}\,[P,\bsigma,\btau]\, u(\bs')\,u(\bs)\,
\delta_{\bs^{(1)},\bsigma'}\,\delta_{\bs^{(2)},\bsigma}}
{\lambda_0(n)\, \sum_{\bs}u(\bs)\,u(\bs)}. \nonumber
\end{eqnarray}
Equations (\ref{eq:right}) and (\ref{eq:P_decomp}) are the basis
of our analysis in the subsequent sections.

\subsection{Replica symmetry and self-consistent equations}
\label{sec:RS}

Since  the order function $P$ depends on $n$ via the
dimensionality of its arguments we must now make an ansatz that
will allow us eventually to take the limit $n\to 0$. The simplest choice
corresponds to considering permutation invariance of $P$ with
respect to  its replica indices. This symmetry is guaranteed by
considering e.g.\@ the following form:
\begin{eqnarray}
P_\mathcal{J}(\bsigma,\btau)=
\int dW_\mathcal{J}(\bh)
\frac{e^{\beta \sum_\alpha (h_1\s^\alpha+h_2\tau^\alpha +h_3\s^\alpha\tau^\alpha)}}{[\mathcal{N}(\bh)]^n},
\label{eq:W_def}
\end{eqnarray}
with the short-hand notation $\bh=(h_1,h_2,h_3)$,
$dW_\mathcal{J}(\bh)=d\bh\,W_\mathcal{J}(\bh)$ and  $\mathcal{N}(\bh)$
ensures that $\int d\bh\,W_\mathcal{J}(\bh)=1$. We also assume that the
eigenvector $u(\bs)$ takes the form
\begin{eqnarray}
u(\bs) = \int d\Phi(\bx)\ \label{eq:Phi_def}
e^{\beta \sum_\alpha(x_1 s^{\alpha,1} + x_2 s^{\alpha,2}+ x_3 s^{\alpha,1} s^{\alpha,2})}.
\end{eqnarray}
With these assumptions we can now proceed further and rewrite the
extremisation problem (\ref{eq:extr}) in terms of the pair of
densities $\Phi$ and $W_{\mathcal{J}}$. The starting point is
equations (\ref{eq:right}) and (\ref{eq:P_decomp}) respectively.
Inserting our assumptions (\ref{eq:W_def}) and (\ref{eq:Phi_def})
leads after some algebra to the following set of closed equations
for $n\to0$:
\begin{eqnarray}
\lefteqn{\lambda_0(0)\,\Phi(\bx')\nonumber}
\\
&& = \sum_{\mu\geq 0}\frac{e^{-c}c^\mu}{\mu!}\Bra\int
[\prod_{\nu=1}^\mu dW_{\mathcal{J}_{\nu}}(\bh_\nu)] \int
d\Phi(\bx)\right.\nonumber
\\
&&
\left.\times\prod_{i=1}^3 \delta\left[x_i'-\frac{1}{4\beta} \sum_{\sigma\tau}f_i(\sigma,\tau)G_{\sigma,\tau}(\bx,\bh_\nu)\right]\Ket_{\mathcal{J}_{1}\cdots\mathcal{J}_{\mu}}
\label{eq:self_Phi}
\end{eqnarray}

\begin{eqnarray}
\lefteqn{\lambda_0(0)\, W_\mathcal{J}(\bh')\nonumber}
\\
&& =\sum_{\mu\geq 0}\frac{e^{-c}c^\mu}{\mu!}\Bra\int
[\prod_{\nu=1}^\mu dW_{\mathcal{J}_\nu}(\bh_\nu)] \int d\Phi(\bx)
d\Phi(\bx')\right.
\nonumber\\
&\times&
\left.\prod_{i=1}^3\delta\left[h_i'-\frac{1}{4\beta}\sum_{\sigma\tau}f_i(\sigma,\tau)
H_{\sigma,\tau}^{(\mathcal{J})}(\bx,\bx',\bh_\nu)\right]
\Ket_{\mathcal{J}_1\cdots\mathcal{J}_\mu}\label{eq:self_W}
\end{eqnarray}
In the above we have used the function $f_i(\sigma,\tau)$ with
\[
f_1(\sigma,\tau) = \sigma,\quad f_2(\sigma,\tau) = \tau,\quad
f_3(\sigma,\tau) = \sigma\tau,
\]
while $G,H$ correspond to
\begin{eqnarray}
\lefteqn{G_{\sigma,\tau}(\bx,\bh)=\nonumber}
\\
&&\hspace{-3mm}
\log\left\{\sum_{\tau_1\cdots\tau_\mu}\sum_{\eta=\pm}
e^{\beta(x_1\tau+\eta x_2+\tau
x_3)}S_{\tau,\sigma,\eta}(\sum_{\nu\leq
\mu}\mathcal{J}_\nu\tau_\nu)\right\}\label{eq:update_G}
\end{eqnarray}
\begin{eqnarray}
H^{(\mathcal{J})}_{\sigma,\tau}(\bx,\bx',\bh)&=&\log\left\{\sum_{\tau_1\cdots\tau_\mu}
\sum_{\eta,\omega=\pm} e^{\beta(x_1\sigma+\omega(x_2+\sigma
x_3))}\right.\nonumber
\\
& &\hspace{-20mm}\times\ \left.e^{\beta(\sigma x_2'+\eta
(x_1'+\sigma x_3'))}S_{\sigma,\omega,\eta}(\sum_{\nu\leq
\mu}\mathcal{J}_\nu\tau_\nu+\mathcal{J}\tau)\right\}\label{eq:update_H}
\end{eqnarray}
and
\begin{eqnarray*}
\lefteqn{S_{\sigma_1,\sigma_2,\sigma_3}(x)=\prod_{\nu\leq
\mu}e^{\beta(h_{1\nu}\tau_\nu+h_{2\nu}\sigma_1+h_{3\nu}\sigma_1\tau_\nu+\frac{1}{2c}\sigma_1
\mathcal{J}_\nu\tau_\nu)}}
\\
& &\times\ e^{\frac12\beta J_0\sigma_1(\sigma_2+\sigma_3)}
\Theta\left(J_0\sigma_1(\sigma_2+\sigma_3)+\frac{\sigma_1}{c}x\right).
\end{eqnarray*}

Finally, to calculate (\ref{eq:extr}) we need to determine the
largest eigenvalue $\lambda_0(n)$ for $n\to 0$. The starting point
here is our eigenvector equation (\ref{eq:right}). Evaluating the
traces over the spin variables with the
definitions of the transfer matrix (\ref{eq:matrix_T}) and
eigenvectors (\ref{eq:Phi_def}) leads for $n\to 0$ to
\begin{eqnarray}
\lefteqn{\lambda_0(n)=1+n\left\{\sum_{\mu\geq
0}\frac{e^{-c}c^\mu}{\mu!}\int
[\prod_{\nu\leq\mu}dW_{\mathcal{J}_\nu}(\bh_{\nu})]\right.\label{eq:lambda}}
\\
&& \hspace{-4mm} \left. \frac{\Bra\int
d\Phi(\bx)\left[\log\left(\frac{K(\bx,\{\bh_\nu\})}{A(\bx)}\right)-\sum_{\nu\leq
\mu}\log\mathcal{N}(\bh_\nu)\right] \Ket_{\{\mathcal{J}_\nu\}}}{\int
d\bx'\Phi(\bx')}
 \right\}
 \nonumber
 \\
 & &
 +\mathcal{O}(n^2), \nonumber
\end{eqnarray}
so that $\lambda_0(0)=1$. We have defined
\begin{eqnarray}
A(\bx) &=& \sum_{\sigma,\sigma'=\pm}e^{\beta( x_1\sigma+
x_2\sigma'+x_3\sigma\sigma')} \label{eq:A}
\\
K(\bx,\{\bh_\nu\}) & = & \sum_{\tau^1\cdots
\tau^\mu}\sum_{\sigma_1,\sigma_2,\sigma_3}e^{\beta (x_1 \s_1+
x_2\sigma_3+ x_3\sigma_1 \s_3)}\nonumber
\\
& & \times S_{\sigma_1,\sigma_2,\sigma_3}(\sum_{\nu\leq
\mu}\mathcal{J}_\nu\tau_\nu)\label{eq:K}.
\end{eqnarray}
With the expression (\ref{eq:lambda}) we can now evaluate
(\ref{eq:extr}):
\begin{eqnarray}
 \lefteqn{-\beta f
= -\frac{c}{2}
\Bra\int dW_{\mathcal{J}}(\bh)\,dW_{\mathcal{J}}(\bh')\log D(\bh,\bh')\Ket_{\mathcal{J}}} \nonumber
\\
& & +\Bra\sum_{\mu\geq 0}\frac{e^{-c}c^\mu}{\mu!}\int
\prod_{\nu\leq \mu} dW_{\mathcal{J}_\nu}(\bh_\nu)\right.\nonumber
\\
& & \times\left.\frac{1} {\int d\bx'\ \Phi(\bx')}\int
d\bx\,\Phi(\bx) \,\log \left[\frac{K(\bx,\{\bh_\nu\})}{ A(\bx)}
\right] \Ket_{\mathcal{J}_1\cdots\mathcal{J}_\mu} \label{eq:f_final}
\end{eqnarray}
with
\[
D(\bh,\bh')=\sum_{\sigma,\sigma'=\pm} e^{\beta( h_1\sigma+
h_2\sigma' +h_3\sigma\sigma'+h_1'\sigma'+h_2'\sigma +
h_3'\sigma\sigma')} \nonumber,
\]
which is our final result. From this equation we can inspect the physical meaning 
of the densities $W_{\mathcal{J}}(\bh)$ and $\Phi(\bx)$. The Poisson distribution $e^{-c}c^\mu/\mu!$ of mean $c$ can be clearly associated to the degree distribution of the graph. Once a degree has been sampled from this distribution one performs $\mu$ integrals over the densities $\{W_{\mathcal{J}_\nu}\}$ and one over $\Phi(\bx)$. Thus, we can think of the $W_{\mathcal{J}_\nu}(\bh)$ as the distribution of `effective' fields (or, so-called `messages') coming from the long-range connections and $\Phi(\bx)$ as those coming from the ring neighborhood.

\subsection{Benchmark tests of the theory}
\label{sec:tests}

Given the complicated structure of our equations, we now wish to
inspect the validity of the theory against simple benchmark tests.

Firstly, in the absence of the `ring' structure, and for
strictly ferromagnetic interactions, the equations must reduce to
those found in Ref.\@ \cite{bergsellito}. Indeed, setting into the
update functions (\ref{eq:update_G},\ref{eq:update_H})   $J_0=0$
and $p=1$ we find that after the first iteration of
(\ref{eq:self_Phi}) the density $\Phi$ collapses to
$\Phi(x_1,x_2,x_3)=\delta(x_1)\phi(x_2)\delta(x_3)$. Details of
the non-trivial function $\phi$ are not important for the purposes
of this section. Filling in this information in the right-hand
side of (\ref{eq:self_W}) leads to several simplifications as a
result of which the dependence of $W_{\mathcal{J}}(\bh)$ on $\phi(x)$ drops out
completely. The resulting closed equation is the one found in
\cite{bergsellito}. Thus, at the level of the self-consistent
equation the expressions reproduce the correct result. Next, we
consider the free energy. Clearly, the energetic part of
(\ref{eq:f_final}) depends only on $W_{\mathcal{J}}(\bh)$ and in the special
benchmark case takes the same form as the energetic term of
\cite{bergsellito}. The entropic term on the other hand, depends
explicitly  on the reduced density of fields $\phi(x)$ which is
coupled to the functions $A(x)$ and $K(x,\bh)$, eq.\@ (\ref{eq:A})
and (\ref{eq:K}) respectively. Here, it turns out that one can
write $K(x,\bh)=A(x)\,\tilde{K}(\bh)$ which effectively removes
$A(x)$ and $\phi(x)$ completely from (\ref{eq:f_final}). The
resulting expression reproduces the free energy of
\cite{bergsellito}.

A second test of the theory is against the small-world
thermodynamic analysis of Ref.\@ \cite{theodore,guzai_rsb}. To map
the generating function (\ref{eq:generating_function}) to the free
energy of that system we set $\Theta(x)=1$ for all $x$. This
removes the stability condition from our definitions. After the
first iteration of (\ref{eq:self_Phi}) we now find that the
function $\Phi(\bx)$ collapses to
$\Phi(x_1,x_2,x_3)=\delta(x_1)\tilde{\phi}(x_2)\delta(x_3-\frac12J_0)$ and
using this to iterate (\ref{eq:self_W}) we obtain that
$W(h_1,h_2,h_3)=w(h_1)\delta(h_2)\delta(h_3-J/{2c})$. Thus, in
both cases only one of the three components is non-trivially
distributed. With these relations we recover at the second
iteration of (\ref{eq:self_Phi},\ref{eq:self_W}) the
self-consistent equations of \cite{guzai_rsb}. Equations (\ref{eq:self_Phi},\ref{eq:self_W}) 
also reduce to (29,30,32) of \cite{theodore} if the analysis of
\cite{theodore} would have been based on symmetric
transfer matrices. In this case the final result
(\ref{eq:f_final}) reduces to the correct free energy.

Finally, by inspection of the physical interpretation of equations
(\ref{eq:self_Phi},\ref{eq:self_W}) and (\ref{eq:f_final}) we can map
our model onto the one of \cite{pagnani} which evaluates the 
number of metastable configurations on a Bethe lattice. This can be done by
appropriately converting the Poisson degree distribution to a Bethe-lattice one.
We have done this test numerically and within the limits of precision we find good agreement with 
the results of \cite{pagnani}.

\section{Results}
\label{sec:results}

We are now interested in obtaining the energy $e$ and entropy densities $s$
of the metastable states. These can be generated from $f$ through
simple relations, i.e.\@ $e=
\partial_\beta (\beta f)$ and $s=\beta\big(e-f\big)$.
To obtain these we have solved equations (\ref{eq:self_Phi}) and
(\ref{eq:self_W}) through `population dynamics'
\cite{mezardparisi} and used simple Monte Carlo integration
recipes to evaluate (\ref{eq:f_final}) (typically, in executing
population dynamics the population size has been taken of the
order of $10^5$ and we assumed algorithmic equilibration after 100
steps). Since the profiles of $W_\mathcal{J}(\bh)$ and $\Phi(\bx)$
depend on the temperature, differentiation of $f$ with respect to
$\beta$ will involve derivatives of both of these densities. One
of these, namely $\partial_W f\cdot\partial_\beta W$, trivially
vanishes as we are at saddle-points of the order function
$P_\mathcal{J}(\bsigma,\btau)$ and, consequently, also of
$W_\mathcal{J}(\bh)$. The derivative $\partial_\Phi f
\cdot\partial_\beta \Phi$ however, may not necessarily vanish as
we have not extremized $f$ with respect to $\Phi$. Therefore, we
cannot assume that the energy $e=
\partial_\beta (\beta f)$ is given by a simple partial differentiation of
(\ref{eq:f_final}) (which indeed leads to incorrect results). To
proceed analytically, one is required to obtain further closed relations
for $\hat{\Phi}=\partial_\beta\Phi$ which, given the complexity of
the equations involved, appears to be a hard task. Here, we have
chosen to carry out the differentiations numerically.
In all cases we have taken the average connectivity to be $c=2$.

\begin{figure}[t]
\begin{center}
\includegraphics[width=.45\textwidth]{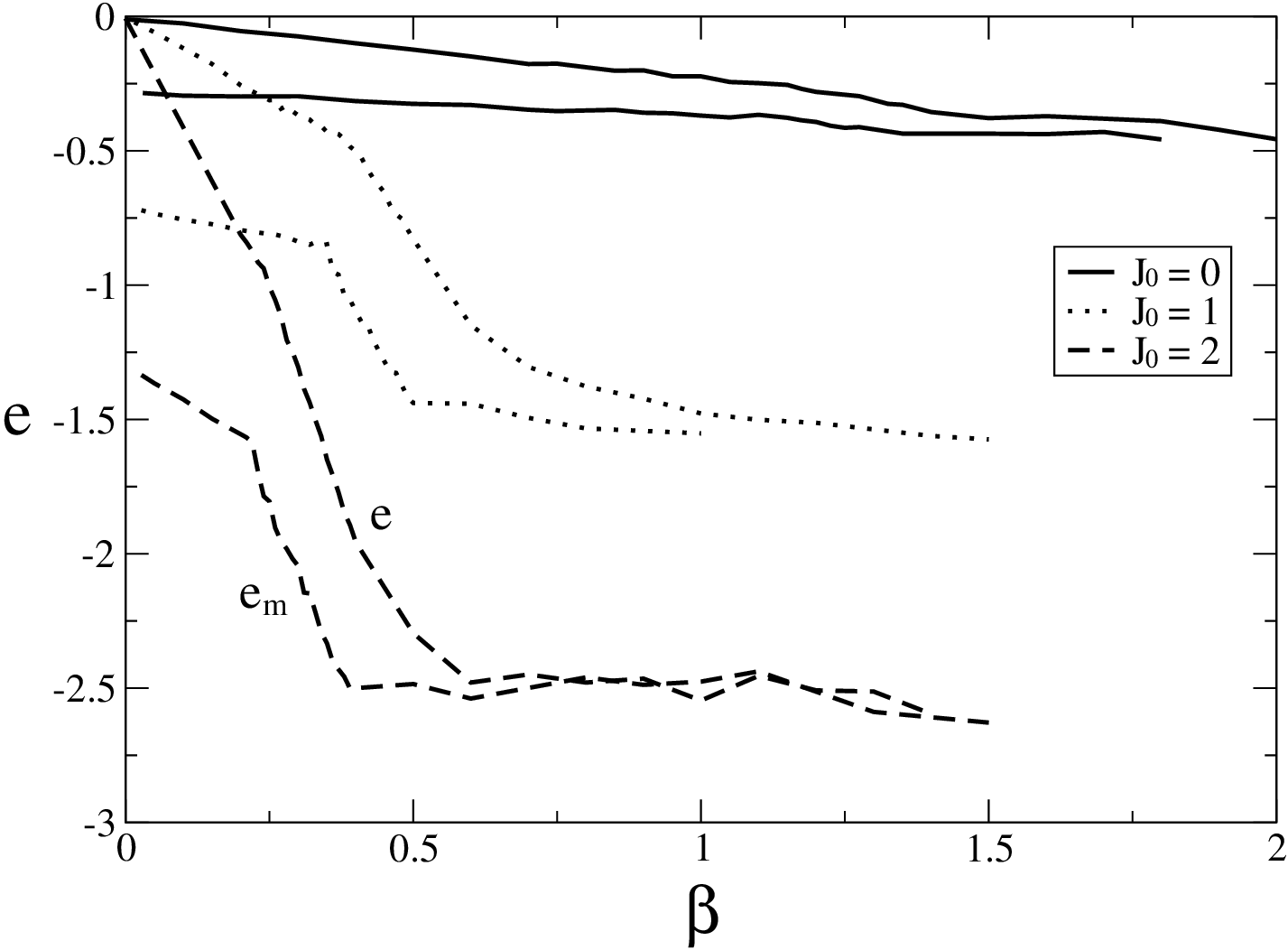}
\caption{Comparison of the energy $e$ of a small-world system in
which spin configurations can visit all possible configurations against
the energy $e_m$ of the metastable configurations, plotted as
function of the inverse temperature $\beta$ and for
$J_0=\{0,1,2\}$. We see that $e_m<e$ and that for $\beta\to\infty$
$e_m\approx e$. Parameter values: $J=1$, $p=1$ and $c=2$.
\label{fig:energy-beta}}
\end{center}
\end{figure}

Let us now describe the results. First we take the simplest case
where long-range interactions are of uniform strength, i.e.\@
$p=1$. Thus, the only source of disorder in the system comes
through the connectivity variables $\{c_{ij}\}$. In figure
\ref{fig:energy-beta} we plot the energy against inverse
temperature for three different values of $J_0$ and with $J=1$. In
each case we compare the energy density of the system $e$ when we
allow all configurations to be visited in phase space
(i.e.\@ with $\Theta(x)=1$ for all $x$) against the energy $e_m$
of only the metastable subset of configurations (with
$\Theta(x)=1$ if $x\geq 0$ and 0 otherwise).

Since, by definition, metastable states are minima of the energy
landscape we expect that $e_m\leq e$ at any given temperature
which indeed is verified by the numerics. As we increase the
strength of the short-range couplings $J_0$ the system will
typically require a higher noise level to destroy the order.  For
this reason we see that the location of the phase transition
towards low energy values decreases with $\beta$ as $J_0$ is
increased. For $\beta\to\infty$ one can find the ground state
energy of the system by simple inspection of the Hamiltonian
(\ref{eq:H}), namely $e_{\rm gr}=-J_0-J/2$. Furthemore, since in
the regime of low temperatures we expect the system to be in a
locally stable state we can anticipate that $e \approx e_m$. These
physical arguments are in agreement with the numerical results of
figure \ref{fig:energy-beta}. On the other hand, for $\beta \to
0$, noise dominates the microscopic spin dynamics and thus the
energy of the system  $e$ typically averages to zero for all
values of $J_0$. We also observe that the transition to the
ordered phase is less smooth for $e_m$ than for $e$. This effect
which has also been reported in \cite{bergsellito} for the special
case of $J_0=0$ is due to non-linearities induced by the Heaviside
function.

\begin{figure}[h]
\vspace{4mm}
\begin{center}
\includegraphics[width=.45\textwidth]{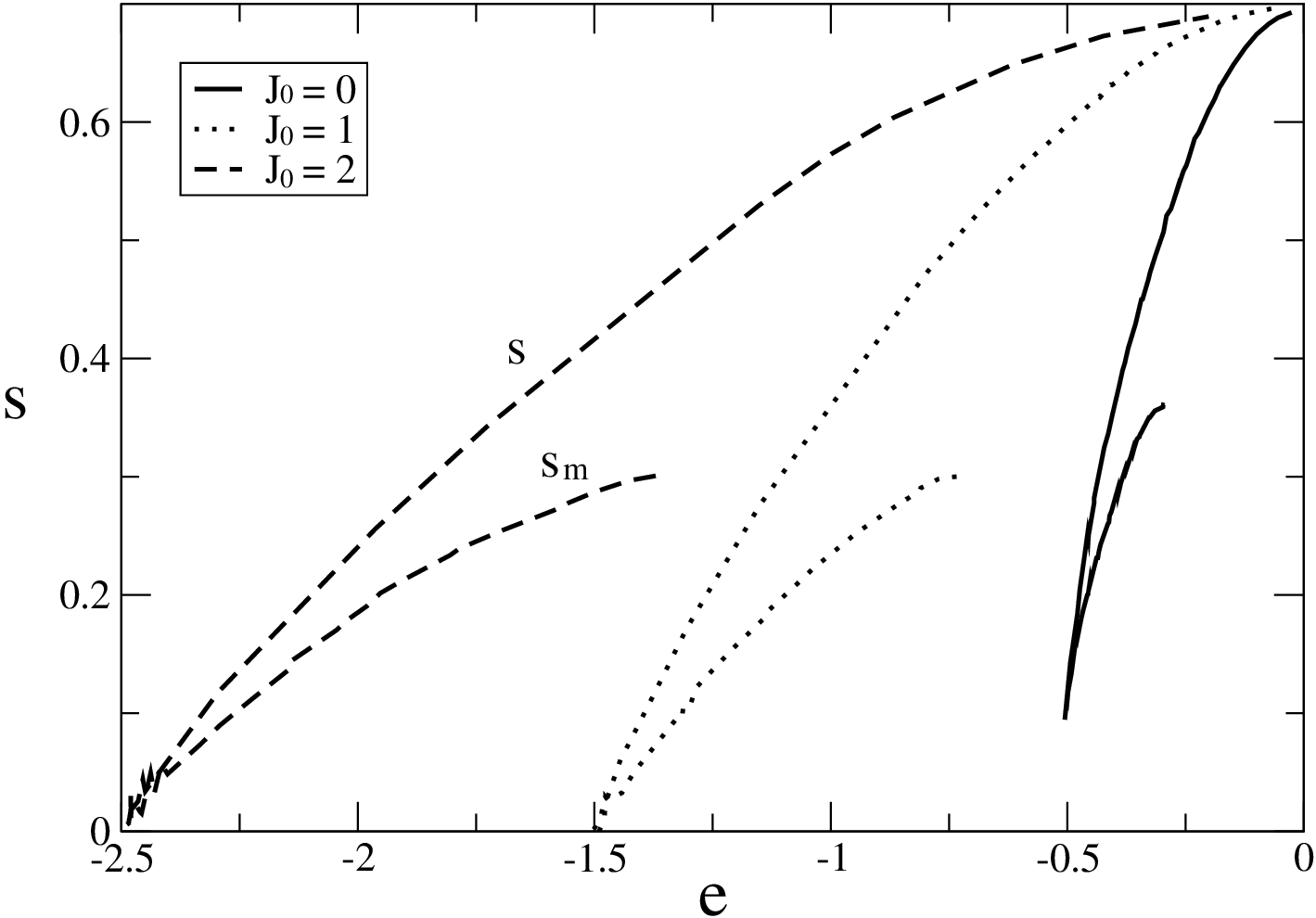}
\caption{Comparison of the entropy $s$ of a small-world system in
which spin configurations can visit all possible configurations against
the entropy $s_m$ of the metastable configurations, plotted
against their energy $e,e_m$, for $J_0=\{0,1,2\}$.
For $J_0=0$ the ground state entropy has a finite value which
vanishes as soon as $J_0>0$. Parameters values: 
$J=1$, $p=1$ and $c=2$.\label{entropy-energy1}}
\end{center}
\end{figure}

In figure \ref{entropy-energy1} we plot the entropy against the
energy for different values of $J_0$, and with $p=1$, $J=1$, $c=2$. The
low-energy part of the graph corresponds to regimes of low
temperatures. As before, we compare the entropy $s$ that would
follow from a thermodynamic calculation in the entire
configuration space against the entropy $s_m$ of the metastable
states. First, we see that one always has $s_m\leq s$, as one
would expect. The difference between the two entropies varies
significantly across the energy axis. For instance, for high
energy values (where the temperature is practically infinite) this
difference reaches its maximum value. On the other hand, for low
temperatures, both entropies reach their minimum value which for
any $J_0> 0$ is zero. In the special case where $J_0=0$ the graph
will typically consist of disconnected clusters which causes the
observed degeneracy. However, as soon as the `ring' connects all
spins together this degeneracy is lost and the ground state
entropy is zero.

\begin{figure}[t]
\begin{center}
\includegraphics[width=.45\textwidth]{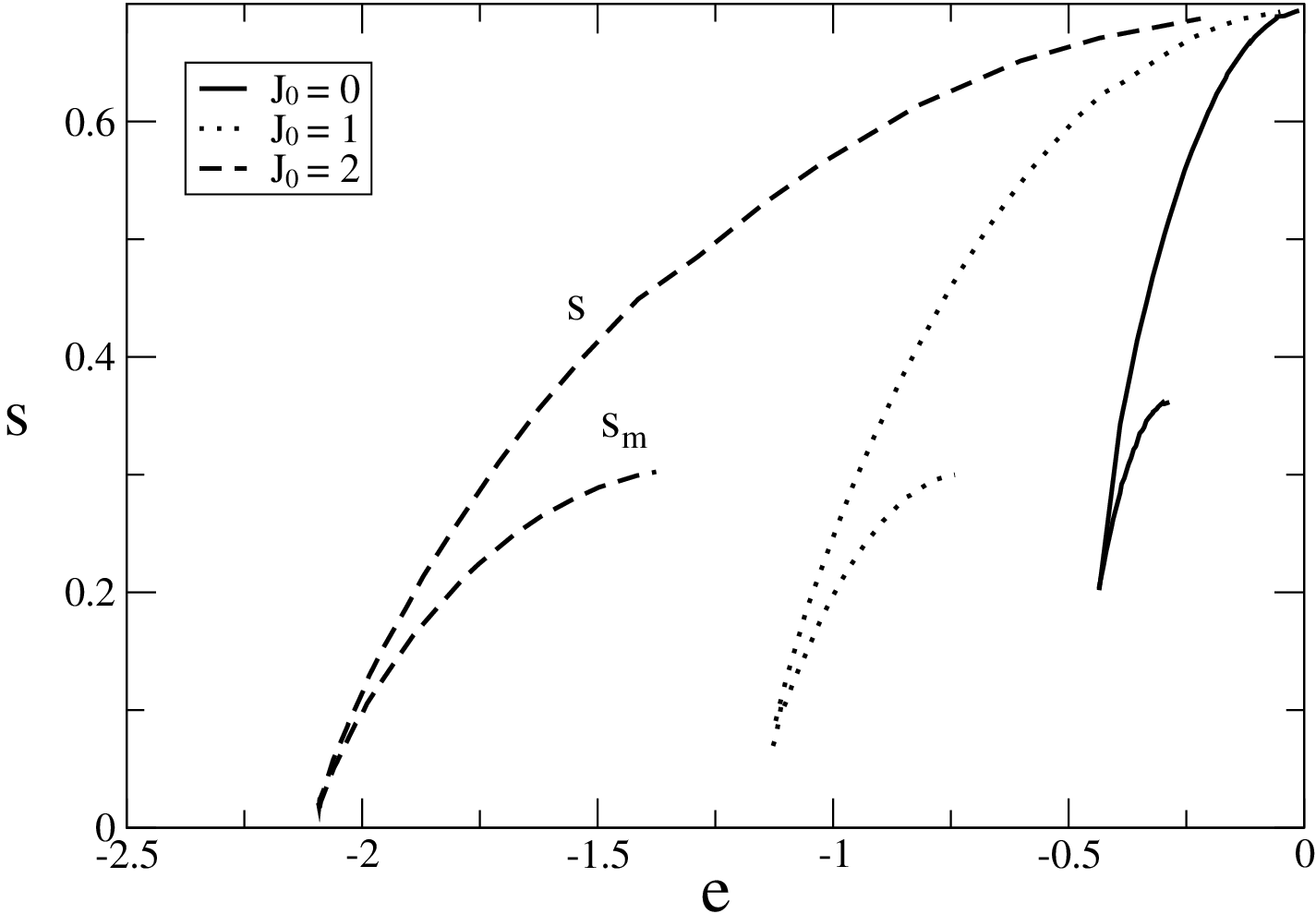}
\caption{We show the entropies $s$ and $s_m$ for small-worlds with
bond-disorder. In this case the degeneracy of the ground state
remains finite even for $J_0>0$. Parameter values: $J=1$, $p=1/2$ and $c=2$.}\label{fig:spinglass2}
\end{center}
\end{figure}

\begin{figure}[t]
\vspace{3mm}
\begin{center}
\includegraphics[width=.45\textwidth]{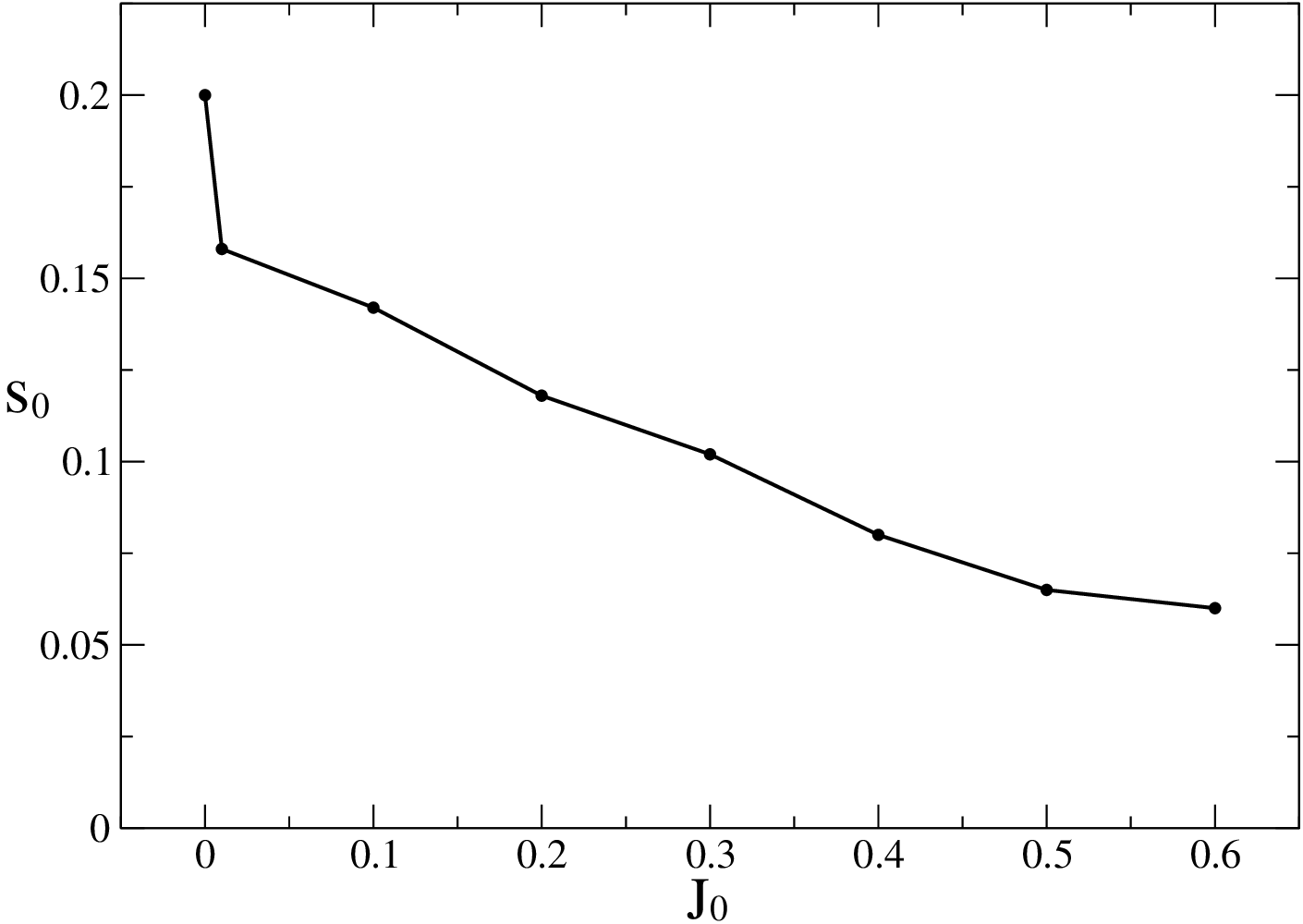}
\caption{The ground state entropy density $s_0$ for a small-world system with
bond-disorder as a function of $J_0$. There is a distinct jump
precisely at $J_0=0$ while the entropy remains finite even for
$J_0>0$. Parameter values: are $J=1$,
$p=1/2$ and $c=2$.} \label{fig:minentropy}
\end{center}
\end{figure}

Let us now examine the case of bond-disorder. In figure
\ref{fig:spinglass2} we present the entropies $s$ and $s_m$
against the energy for different values of $J_0$ and with $J=1$,
$c=2$, $p=1/2$. Firstly, we observe that the ground state energy
is significantly higher compared to the case of $p=1$. This is due
to the value of the local fields which will on average be smaller
for $p<1$ than for $p=1$. We also observe that the ground state
entropy can take a non-zero finite value even at $J_0>0$. This is
due to the presence of anti-ferromagnetic couplings in the system
which increases the fraction of sites with a zero local field.
However, as one increases the strength of the (ferromagnetic)
short-range couplings this fraction of sites becomes smaller and
the degeneracy of the ground state gradually disappears. To
illustrate this effect we plot in figure \ref{fig:minentropy} the
ground state entropy against the short-range coupling strength
$J_0$. We also see a `jump' precisely at $J_0=0$.

\section{Discussion}
\label{sec:conclusions}

In recent years, the theory of complex networks has witnessed a
remarkable growth. Within the area of complex systems, the special
subset of `small-worlds', has aroused the curiosity of theorists
and experimentalists alike due to the striking cooperativity
phenomena that they allow. In particular, for any value of the
average long-range connectivity (however small), small-world
networks can have a phase transition to an ordered phase at a
finite temperature. Small world architectures have been observed
in a wide range of real complex systems.

For a theorist, several important questions arise regarding the
emergent collective properties on such systems. In this paper we
have evaluated the number of metastable configurations. In a
spirit similar to \cite{gardner,bergsellito,pagnani} the `metastability' condition
constrains the partition sum over configurations in which spins
align to their local fields. From an analytic point of view, two
are the main stumbling blocks: firstly, the non-linear nature of
the stability condition, and secondly, the diagonalisation of the
relevant transfer matrix. Our numerical results suggest that, for
low temperatures and in the case of bond-disorder, the metastable
configurations tend to dominate the space of equilibrium states.
We also see that superimposing the one-dimensional `backbone'
structure leads to a significantly smaller degeneracy of the
ground state (which, in fact, vanishes for strictly ferromagnetic
couplings). As function of the short-range coupling $J_0$, there
is a jump in the ground state entropy exactly at $J_0=0$
which is due to the formation of disconnected clusters within the
graph.

\begin{acknowledgements}
We would like to thank J.\@ Berg, T.\@ Coolen, J.\@ Hatchett, I.\@
P\'erez Castillo, A.\@ Pagnani and G.\@ Semerjian for insightful
discussions. This work is partially funded by the Fund for
Scientific Research Flanders-Belgium.
\end{acknowledgements}



\appendix

\section{Derivation of the saddle-point expression (\ref{eq:extr})}
\label{sec:ap1}

Our starting point are equations (\ref{eq:f_inter}-\ref{eq:orderfunction_def}). We replace the delta function in (\ref{eq:orderfunction_def}) by its integral representation which results in
\begin{widetext}
\begin{eqnarray}
\lefteqn{ \Bra Z^n\Ket_{\bc,\bJ}\equiv\Bra 
\sum_{\bsigma^1\cdots\bsigma^n}e^{-\beta\sum_\alpha
\mathcal{H}(\bsigma^\alpha)}\prod_{i,\alpha}\Theta(\s_i^\alpha h_i(\bsigma^\alpha))\Ket_{\bc,\bJ}}
\nonumber \\
 &  &
=\int[\prod_{\bsigma,\btau,\mathcal{J}}dP_{\mathcal{J}}(\bsigma,\btau)
d\hat{P}_{\mathcal{J}}(\bsigma,\btau)]\exp\left[
i\sum_{\bsigma,\btau}\sum_{\mathcal{J}}\hat{P}_{\mathcal{J}}
(\bsigma,\btau)P_{\mathcal{J}}(\bsigma,\btau)+\frac{cN}{2}\left(\sum_{\bsigma\btau}\Bra
P_{\mathcal{J}}(\bsigma,\btau)P_{\mathcal{J}}(\btau,\bsigma)\Ket_{\mathcal{J}}\right)\right]
\nonumber
\\
& &
\times\ \sum_{\bsigma^1\cdots\bsigma^n}
\int\{dh_i^\alpha\}\prod_\alpha e^{\frac12\beta\s_i^\alpha h_i^\alpha}\Theta(\sigma_i^\alpha h_i^\alpha)
\nonumber
\\
& & \times\
\prod_{i=1}^N\left\{\int\frac{\{d\hat{h}_i^\alpha\}}{(2\pi)^{n}}
e^{i\sum_{i,\alpha}
\hat{h}_i^\alpha(h_i^\alpha-J_0(\s_{i+1}^\alpha+\s_{i-1}^\alpha))}
\sum_{\mu\geq
0}\frac{1}{\mu!}\sum_{\btau_1\cdots\btau_\mu}\sum_{\mathcal{J}_1\cdots\mathcal{J}_\mu}
\left(\prod_{\nu\leq \mu}
-i\hat{P}_{\mathcal{J}_\nu}(\bsigma_i,\btau_\nu)\,e^{-i\frac{\mathcal{J}_\nu}{c}\sum_{\alpha}\hat{h}_i^\alpha
\tau_\nu^\alpha}\right)\right\}. \nonumber
\end{eqnarray}
\end{widetext}
We now see that the last line of the above expression has factorised over site indices and the integral
over the variables $\{\hat{h}_i^\alpha\}$ can be done immediately. The result is a delta function
which we use to eliminate $\{h_i^\alpha\}$:
\begin{widetext}
\begin{eqnarray}
\Bra Z^n\Ket_{\bc,\bJ} & = &
\int[\prod_{\bsigma,\btau,\mathcal{J}}dP_{\mathcal{J}}(\bsigma,\btau)
d\hat{P}_{\mathcal{J}}(\bsigma,\btau)]\exp\left[
i\sum_{\bsigma,\btau}\sum_{\mathcal{J}}\hat{P}_{\mathcal{J}}
(\bsigma,\btau)P_{\mathcal{J}}(\bsigma,\btau)+\frac{cN}{2}\left(\sum_{\bsigma\btau}\Bra
P_{\mathcal{J}}(\bsigma,\btau)P_{\mathcal{J}}(\btau,\bsigma)\Ket_{\mathcal{J}}\right)\right]
\nonumber
\\
& & \times\ \sum_{\bsigma^1\cdots\bsigma^n}\prod_{i=1}^N
\left\{\sum_{\mu\geq
0}\frac{(-i)^\mu}{\mu!}\sum_{\btau_1\cdots\btau_\mu}
\sum_{\mathcal{J}_1\cdots\mathcal{J}_\mu} \prod_{\nu\leq
\mu}[\hat{P}_{\mathcal{J}_\nu}(\bsigma_i,\btau_\nu)]
\nonumber \right.
\\
& &
\times \left. \int\{dh_i^\alpha\}\prod_\alpha e^{\frac12\beta\s_i^\alpha h_i^\alpha}\Theta(\sigma_i^\alpha h_i^\alpha)\
\delta\left[h_i^\alpha-J_0(\sigma_{i+1}^\alpha+\sigma_{i-1}^\alpha)-\frac{1}{c}\sum_{\nu=1}
\mathcal{J}_\nu\tau_\nu^\alpha\right]\right\}
\nonumber
\\
&=&
\int[\prod_{\bsigma,\btau,\mathcal{J}}dP_{\mathcal{J}}(\bsigma,\btau)
d\hat{P}_{\mathcal{J}}(\bsigma,\btau)]\exp\left[
i\sum_{\bsigma,\btau}\sum_{\mathcal{J}}\hat{P}_{\mathcal{J}}
(\bsigma,\btau)P_{\mathcal{J}}(\bsigma,\btau)+\frac{cN}{2}\left(\sum_{\bsigma\btau}\Bra
P_{\mathcal{J}}(\bsigma,\btau)P_{\mathcal{J}}(\btau,\bsigma)\Ket_{\mathcal{J}}\right)\right]
\nonumber
\\
& & \times\ \sum_{\bsigma^1\cdots\bsigma^n}\prod_{i=1}^N
\left\{\sum_{\mu\geq
0}\frac{(-i)^\mu}{\mu!}\sum_{\btau_1\cdots\btau_\mu}\sum_{\mathcal{J}_1\cdots\mathcal{J}_\mu}
\prod_{\nu\leq \mu}[\hat{P}_{\mathcal{J}_\nu}(\bsigma_i,\btau_\nu)] \nonumber
\right.
\\
& & \times\ \left.
\prod_{\alpha}e^{\frac12\beta\sigma_i^\alpha(J_0(\sigma_{i+1}^\alpha+\sigma_{i-1}^\alpha)-
\frac{1}{c}\sum_{\nu=1}\mathcal{J}_\nu\tau_\nu^\alpha)}\
\Theta\left(\sigma_i^\alpha(J_0(\sigma_{i+1}^\alpha+\sigma_{i-1}^\alpha)+\frac{1}{c}\sum_{\nu=1}
\mathcal{J}_\nu\tau_\nu^\alpha)\right)\right\}
\end{eqnarray}
\end{widetext}
With  the above, the generating function (\ref{eq:generating_function}) can be written as an extremisation problem, namely:
\begin{widetext}
\begin{equation}
-\beta f= \lim_{n\to 0}\frac{1}{n}\underset{P,\hat{P}}{\rm Extr}
\left[i\sum_{\bsigma,\btau}\sum_{\mathcal{J}}\hat{P}_{\mathcal{J}}
(\bsigma,\btau)P_{\mathcal{J}}(\bsigma,\btau)+\frac{c}{2}\sum_{\bsigma\btau}\Bra
P_\mathcal{J}(\bsigma,\btau)P_\mathcal{J}(\btau,\bsigma)\Ket_\mathcal{J}
+\lim_{N\to\infty}\frac1N\log\,{\rm tr}\,(\tilde{T}^N[\hat{P}])\right]
\label{eq:extr_app}
\end{equation}
\end{widetext}
with the transfer function $\tilde{T}[\hat{P}]$ given by
\begin{widetext}
\begin{eqnarray}
\tilde{T}_{\bsigma_{i-1},\bsigma_i,\bsigma_{i+1}}[\hat{P}]& = &
\sum_{\mu\geq 0}\frac{(-i)^\mu}{\mu!}
\sum_{\btau_1\cdots\btau_\mu}\sum_{\mathcal{J}_1\cdots\mathcal{J}_\mu}
\prod_{\nu\leq
\mu}[\hat{P}_{\mathcal{J}_\nu}(\bsigma_i,\btau_\nu)]\nonumber
\\
& &\times\
\prod_{\alpha}e^{\frac12\beta\sigma_i^\alpha(J_0(\sigma_{i+1}^\alpha+\sigma_{i-1}^\alpha)-
\frac{1}{c}\sum_{\nu=1}\mathcal{J}_\nu\tau_\nu^\alpha)}\
\Theta\left(\sigma_i^\alpha(J_0(\sigma_{i+1}^\alpha+\sigma_{i-1}^\alpha)+\frac{1}{c}\sum_{\nu=1}
\mathcal{J}_\nu\tau_\nu^\alpha)\right)
\end{eqnarray}
\end{widetext}
Variation of (\ref{eq:extr_app}) with respect to the function $P_{\mathcal{J}}$ gives the relation $\hat{P}_{\mathcal{J}}(\bsigma,\btau)=icQ_J(\mathcal{J})\,P_{\mathcal{J}}(\btau,\bsigma)$ for $\mathcal{J}=\{-J,J\}$. Using this identity in (\ref{eq:extr_app}) leads to (\ref{eq:extr}).

\pagebreak

\end{document}